\documentclass[cpp,a4paper,fleqn]{w-art}

\usepackage{times,cite,w-thm}

\usepackage{color}

\usepackage{graphicx}
\usepackage{dcolumn}
\usepackage{bm}
\usepackage{subfigure}
\usepackage{amssymb}
\usepackage{multirow}
\graphicspath{{plots/}}

\def\be{\begin{equation}}
\def\ee{\end{equation}}
\def\bea{\begin{eqnarray}}
\def\eea{\end{eqnarray}}

\newcommand{\bfr}{{\bf r}}

\newcommand{\bfu}{{\bf u}}

\begin{document}

\title{Quantum hydrodynamics for plasmas -- \\ a Thomas-Fermi theory perspective}

\author[D. Michta {\em et al.}]{D.~Michta\inst{1,2}}
\author[]{F.~Graziani\inst{1}
  \footnote{Corresponding author\quad E-mail:~\textsf{graziani1@llnl.gov},
}}
\author[]{M. Bonitz\inst{3}}

\address[\inst{1}]{Lawrence Livermore National Laboratory, Livermore, CA, 94550, USA}

\address[\inst{2}]{Princeton Plasma Physics
Laboratory, Princeton, NJ, 08540, USA}

\address[\inst{3}]{Institute for Theoretical Physics and Astrophysics, Christian Albrechts University Kiel, \\ Leibnizstrasse 15, D-24098 Kiel, Germany}

\date{\today}

\begin{abstract}
The idea to describe quantum systems within a hydrodynamic framework (quantum hydrodynamics, QHD) goes back to Madelung and Bohm. While such a description is formally exact for a single particle, more recently the concept has been applied to many-particle systems
by Manfredi and Haas [Phys. Rev. B {\bf 64}, 075316 (2001)]
and received high popularity in parts of the quantum plasma community. Thereby, often the applicability limits of these equations are ignored, giving rise to unphysical predictions. 
Here we demonstrate that modified QHD equations for plasmas can be derived from Thomas-Fermi theory including gradient corrections. This puts QHD on firm grounds. At the same time this derivation yields a different prefactor, $\gamma=(D-2/3D)$, in front of the quantum (Bohm) potential which depends on the system dimensionality $D$. Our approach allows one to identify the limitations of QHD and to outline systematic improvements.
\end{abstract}

\maketitle

\section{Introduction}\label{s:intro}
Dense plasmas containing quantum degenerate electrons are presently of increasing interest in many fields including condensed matter, astrophysics, warm dense matter, high energy density matter and laser plasmas, e.g. \cite{horing_cpp11,kremp-etal.pre99, graziani11,militzer12}.
Despite of the experimental relevance of these systems, the theoretical desciription of quantum plasmas including long range Coulomb interaction remains far from completed, and a large variety of different theoretical concepts is being applied, including quantum Monte Carlo methods \cite{ceperley, filinov00, militzer12, schoof11} and density functional theory, e.g. \cite{xu, redmer, desiarlais, clerouin92, bonitz_pre_comm} and references therein.
 The situation is even more critical in the area of nonequilibrium plasma behavior.
There are two main lines of research: The first are many-body theory approaches such as quantum kinetic theory and nonequilibrium Green functions methods, e.g. \cite{kremp-book,bonitz-book,bonitz-aip12,bonitz_cpp13}, linear response theory, e.g. \cite{roepke, zhandos_pre15} and time-dependent density functional theory. The second line comprises particle-based computer simulations -- including semiclassical molecular dynamics with quantum potentials, e.g. \cite{afilinov_jpa03,afilinov_pre04,graziani11} or various versions of quantum molecular dynamics, e.g. \cite{desiarlais}. 

These simulations are often very time consuming, so it is tempting to develop simpler schemes that avoid the treatment of the momentum dependence entirely and resort to a  much simpler fluid description. Fluid approaches have proven very successful in classical physics, and it is well known how to include inter-particle correlations, e.g. \cite{dubin96, kaehlert10, dufty13}. Therefore, similar approaches are of high interest for quantum systems as well.
In fact, it has already been shown by Madelung \cite{madelung} and Bohm \cite{bohm} that the Schr\"odinger equation for a single particle can also be transformed to a fluid-like form, see \cite{springer-buch, khan-bonitz14} for an overview. Extension to quantum many-particle systems date far back.
Quantum hydrodynamic models for superfluid bosons were presented, among others, by Gross \cite{gross61,gross63}, Pitaevski \cite{pitaevski61} and Biswas \cite{biswas75}. Finally, the extension 
to plasmas---i.e. many-fermion systems---has been performed by Manfredi and Haas (MH) in Ref.~\cite{manfredi-haas01}. Similar equations have been derived by Kuzmenkov and co-workers \cite{kuzmenkov00, kuzmenkov01}. Assuming factorization of the $N$-particle wave function into equally occupied single-particle orbitals MH derived a simple set of fluid equations very similar to the one-particle case and to the case of bosons. Although these appear to be rather drastic assumptions, 
these quantum hydrodynamics (QHD) equations  have been used in recent years in a very large number of works in quantum plasmas, for the computation of linear and nonlinear oscillations and waves. Applications include warm dense matter, dense astrophysical plasmas, electrons in metals and electron-hole plasmas in semiconductors. Furthermore, the QHD equations have been extended to relativistic plasmas as well as to include the spin density as an independent variable, e.g. \cite{marklund07}. We also mention attempts to approximately include exchange and correlation effects \cite{manfredi08}.

A crucial question is the range of validity of QHD. Unfortunately, many recent applications have ignored the limitations, giving rise to predictions that are in conflict with basic results of many-body quantum physics. As a consequence some controversy about this popular theory has emerged that concerned, among others, the prediction of an attractive proton-proton interaction in dense zero temperature quantum plasmas \cite{shukla_prl,bonitz_pre_comm, bonitz_pscr, stanton_pre_15}
and the prediction of giant spin polarizations and spin-driven lasing \cite{asenjo11,braun12, krishnaswami14, krishnaswami15}. Among more general criticisms we note the papers by Vranies et al. \cite{vranies,vranies_re} and Vladimirov et al. \cite{vladimirov} 
pointing out additional limitations. Thus a reliable answer about the quality and limitations of the QHD predictions is important for the field of quantum plasmas.

In this paper we approach this issue from an entirely different side. We use, as the starting point, an independent
 well established continuum approach in quantum many-fermion physics: Thomas-Fermi theory (TF). We show that the QHD equations can be derived from TF  with gradient corrections. While this gives strong conceptional support for QHD, the resulting equations have a different explicit form than those proposed by MH: they differ in the prefactor of the Bohm term. This correspondence between QHD and TF allows for the derivation of improved approximations for the former based on the extensive experience accumulated in the latter.

\section{Summary of Quantum hydrodynamics equations}\label{s:qhd}
The QHD equations derived by MH \cite{manfredi-haas01} consist of the continuity equation and the momentum density balance
\begin{align}
 \partial_t n + \nabla_R (n{\bf u}) &= 0,
\label{eq:continuity_n}\\
 mn\left(\partial_t + {\bf u}\nabla_R \right) {\bf u} &=
-n \nabla_{R} V - \nabla_{R}\left( p^{\rm MB} + \pi  \right),
\label{eq:current_balance_n}
\end{align}
where $V$ is a single-particle potential energy including external and induced electrostatic (Hartree) contributions, $V=V_{\rm ext}+V_{\rm H}$.
The closure of the inifinite system of fluid equations is made by choosing the many-body pressure $p^{\rm MB}$ as the one of an 
ideal Fermi gas at $T=0$, assuming locality in space and time (adiabatic local density approximation, ALDA):
\begin{align}
p^{\rm MB}({\bf r},t) = \frac{2}{5}n_0E_F(n_0)\left( \frac{n({\bf r},t)}{n_0} \right)^{5/3},
\label{eq:p_mb} 
\end{align}
where $E_F$ is the Fermi energy and $n_0$ is a  reference density. This is analogous to classical hydrodynamics 
where $p^{\rm MB} = nk_BT$. The additional pressure $\pi$ is a pure quantum effect resulting from the quantum kinetic energy 
(finite extension of quantum particles) and exists even for a single particle \cite{madelung, bohm}. It is related to the 
Bohm potential $V_B$,
\begin{align}
\nabla \pi = n\nabla V_B, \qquad V_B = \frac{\hbar^2}{2m}\frac{\nabla^2 \sqrt{n}}{\sqrt{n}}, \qquad n=A^2,
\label{eq:p_bohm} 
\end{align}
where $A$ is the (real) amplitude of the wavefunction. While this system (with $p^{\rm MB}=0$) is exact for a single quantum 
particle, its extension to a many-particle system \cite{manfredi-haas01} is an approximation. The first is the assumption of a spatial averaging (coarse graining) over length scales of (at least) a few interparticle distances ${\bar r}$ which is typcial for any fluid approach \cite{bonitz_pre_comm, bonitz_pscr}.
Furthermore, aside from postulating Eq.~(\ref{eq:p_mb}), it is 
assumed that the system has no correlations. For a plasma this means that the coupling strength should be negligible, i.e. 
the ratio of the mean Coulomb energy to the kinetic energy, $\langle e^2/r \rangle \ll \langle E_{kin}\rangle$. In the ground state 
this amounts to $\langle e^2/r \rangle \ll E_F$ [where $E_F$ is the Fermi energy] or $r_s = {\bar r}/a_B \ll 1$ [$a_B$ is the Bohr radius], e.g. \cite{bonitz_pre_comm, khan-bonitz14}. 
Finally, to obtain Eq.~(\ref{eq:p_bohm}), exchange effects are neglected; so the $N$-particle wave function of the non-interacting system factorizes, 
\begin{align}
\Psi_N({\bf r}_1,\dots {\bf r}_N)=\Pi_{i=1}^N A_i({\bf r}_i,t)e^{-iS_i/\hbar},
\label{eq:factorization} 
\end{align}
where $A_i$ and $S_i$ are real functions and, furthermore, all amplitudes are assumed equal \cite{manfredi-haas01}, $A_i({\bf r},t) = A({\bf r},t)$. The combination of these approximations makes it very difficult to assess the quality of the solutions of the QHD equations, 
for a recent analysis, see Refs.~\cite{manfredi08, khan-bonitz14} and references therein. 

\section{Thomas-Fermi theory approach to QHD}\label{s:tf}
A different approach to QHD is possible starting from a suitable energy or action functional of the plasma and performing a variational minimization. Such ideas have been used in many fields. For example, Bloch \cite{bloch} developed a hydrodynamical theory of the electron gas. Gross \cite{gross61} and Pitaevski \cite{pitaevski61} derived QHD equations for a weakly interacting superfluid Bose condensate -- the famous Gross-Pitaevski equation. The connection between a quantized field theory and hydrodynamic equations was given in Ref.~\cite{gross63}. A variational principle for the Gross-Pitaevski equation, based on a time-dependent Lagrange density, was studied in Ref.~\cite{perez_96}. For an overview containing QHD-type equations for Bose systems, see Ref.~\cite{pfau_09}.

\subsection{Variational formulation}
Starting point is the total energy of a one-component quantum system consisting of  a mean field (Hartree, ``H'') contribution, the energy in an external potential (``ext''), kinetic energy and a residual ``exchange-correlation (xc)'' term, written as a functional of density,
\begin{align}
 E[n] &= U_{\rm ext}[n] + U_{\rm H}[n] + T[n] + U_{\rm xc}[n],
\label{eq:e-funct}
\\
U_{\rm ext}[n] &= \int d^3r \,V_{\rm ext}({\bf r})\,n({\bf r}), \quad U_{\rm H}[n] = \frac{1}{2}\int d^3r \,V_{\rm H}([n],{\bf r})\,n({\bf r}).
\nonumber
\end{align}
The ground state density profile $n_0({\bf r})$ is found by minimizing this functional, under the constraint of a given total particle number $N=\int d^3r \,n({\bf r})$,
\begin{align}
 0 = \delta\left[ E - \mu \int d^3r \,n({\bf r}) \right],
\nonumber
\end{align}
with the variation yielding the chemical potential
\begin{align}
 \mu[n_0] &=  V_{\rm ext} +V_{\rm H}[n_0] + \frac{\delta T[n]}{\delta n}\bigg|_{n=n_0} + \frac{\delta U_{\rm xc}[n]}{\delta n}\bigg|_{n=n_0}. 
\label{eq:mu}
\end{align}
Stability of the ground state profile is associated with vanishing of the total force, ${\bf F}[n_0] = -{\bf \nabla}\mu[n_0]=0$.

The transition to hydrodynamics is straightword by allowing for perturbations of the density around the ground state profile and for currents (not necessarily small) that are driven by the selfconsistent force 
${\bf F}[n({\bf r},t)]=-{\bf \nabla}\mu[n({\bf r},t)]$. Within the adiabatic LDA it is assumed that the functional form of $\mu$, Eq.~(\ref{eq:mu}), remains the same:
\bea
\nonumber
\frac{\partial}{\partial t}n(\bfr,t) + \nabla\cdot\left[ n(\bfr,t)\bfu(\bfr,t)\right] &=& 0,\\
\label{qhd}
m \frac{\partial }{\partial t}\bfu(\bfr,t) + m\bfu(\bfr,t) \cdot\nabla \bfu(\bfr,t) &=& 
- {\bf \nabla}\mu[ n(\bfr,t)].
\eea
This hydrodynamic set of equations includes, selfconsistently via (\ref{eq:mu}), all forces arising from the kinetic, Hartree, external, and exchange-correlation energy. 
It remains to choose appropriate expressions for the ideal part (kinetic energy) and the exchange-correlation term. Here we concentrate on the kinetic energy and neglect $U_{\rm XC}$, since this is beyond standard QHD.

\subsection{Gradient corrections. Recovery of the Bohm potential of QHD}\label{s:gradient}
At zero temperature the kinetic energy $T$ can be decomposed, using a systematic gradient expansion around the homogeneous limit,
\be
\label{TFGC}
T[n] = T_{\rm TF}[n] + T_{\rm GC}[n].
\ee
Here $T_{\rm TF}[n]$ is the Thomas-Fermi functional (local part), and $T_{\rm GC}[n]$ contains the gradient corrections,
with the result 
\begin{align}
 T[n] &= \int d^3r \,\left\{ t[n] \,n({\bf r}) + \gamma\,\frac{\hbar^2}{8m} \frac{|\nabla n|^2}{n} + O[(\nabla n)^3] \right\},
\label{eq:t}
\end{align}
where the ideal local energy density is $t[n]= \frac{3}{5}E_F[n] \sim n^{2/3}$. The gradient term was first computed by 
von Weizs\"acker \cite{weizsaecker} for fermions and bosons in the ground state who obtained the expression (\ref{eq:t}) with $\gamma=1$. While his result is correct for 
bosons, his calculation for fermions is not, as we briefly discuss in Sec.~\ref{s:w-mh}. The  fermion problem was subsequently analyzed
by Kompaneets \cite{kompaneets56} and many others, see e.g. Ref. \cite{kirzhnits75}, for an overview. The correct result of the gradient expansion
 for fermions is due to Kirzhnitz \cite{kirzhnits} who performed a gradient expansion of the one-particle Green function and 
obtained expression (\ref{eq:t}) with the coefficient $\gamma=1/9$.

Let us now establish the connection between the kinetic energy functional (\ref{eq:t}) and the QHD equations (\ref{qhd}). To this end we compute the functional derivative on the r.h.s. of the momentum balance (\ref{qhd}), with the result
\begin{align}
 \frac{\delta T_{\rm TF}[n]}{\delta n} &= E_F[n],
\\
 \frac{\delta T_{\rm GC}[n;\gamma]}{\delta n} &= \gamma\, \frac{\hbar^2}{8m}\left( \left| \frac{\nabla n}{n} \right|^2 - 2 \frac{\nabla^2 n}{n} \right) 
 = \gamma \,V_B[n].
\label{eq:grad=vb}
\end{align}
Thus, we have established a direct connection between the energy functionals of Thomas-Fermi theory with gradient corrections and quantum hydrodynamics. Let us summarize the main conclusions following from (\ref{eq:grad=vb}).
\begin{enumerate}
 \item The functional derivative of the Thomas-Fermi term yields exactly the Fermi pressure $p^{\rm MB}$, Eq.~(\ref{eq:p_mb}), justifying the closure 
approximation of the QHD equations. 
 \item The functional derivative of the gradient correction coincides, up to the coefficient $\gamma$, with the Bohm potential (\ref{eq:p_bohm}) of QHD.
 \item The Bohm potential, $V_B$, is directly connected with the von Weizs\"acker functional, (i.e. with $T_{\rm GC}[\gamma=1]$). Since the first order gradient expansion result $\gamma=1$ is exact for bosons, the QHD equations in the form (\ref{eq:current_balance_n}) with the Bohm potential (\ref{eq:p_bohm}) apply to (non-interacting) bosons, within the validity limits of a hydrodynamic description. In fact, these QHD equations are equivalent to the time-dependent Gross-Pitaevski equation and are widely used in the field of ultracold bosonic atoms, e.g. \cite{pfau_09}.
 \item Another case where the Bohm potential is equivalent to the von  Weizs\"acker gradient correction result ($T_{\rm GC}[\gamma=1]$) is the one  of distinguishable particles. An example are electrons that are tightly bound in atoms such that their wave function overlap is negligible. 
 This allows for an efficient quantum trajectory-based description of chemical dynamics \cite{bittner}.
 \item The only case where the von  Weizs\"acker gradient correction result ($T_{\rm GC}[\gamma=1]$) is (at least formally) correct for fermions is the case of a single particle or 
 two non-interacting particles with opposite spin projection. In the former case the density $n$ in Eq.~(\ref{eq:t}) is just the modulus squared of the wave function. In the latter case the density is the sum of the contributions from both particles.
 \item The gradient expansion for fermions has been extended to systems of different dimensionality $D$ by Holas {\em et al.} \cite{holas91} and Salasnich \cite{salasnich07}, who recovered the functional $T_{\rm GC}[\gamma]$, but with the $D$-dependent value $\gamma=\frac{D-2}{3D}$. While for 3D systems this agrees with the result of Kirzhnitz ($\gamma^{\rm 3D}=1/9$), for 1D systems one obtains, instead, $\gamma^{\rm 1D}=-1/3$. For 2D systems the coefficient is zero meaning that the gradient expansion has to be extended to higher orders, as was done e.g. by Engel {\em et al.} \cite{engel89}.
\end{enumerate}

\subsection{Comparison of the approaches of von Weizs\"acker and Manfredi-Haas}\label{s:w-mh}
The exact agreement of the Bohm potential $V_B$ in the QHD equations with the von Weizs\"acker gradient correction ($T_{\rm GC}[\gamma=1]$) is striking and makes a more detailed comparison of the two approaches and their respective assumptions interesting. In his remarkable paper \cite{weizsaecker} von Weizs\"acker 
considered a system of many noninteracting fermions (protons and neutrons in nuclei) and attempted to generalize the local Thomas-Fermi theory by the inclusion of gradient corrections.
He used the occupation number representation, thereby properly assuring the Pauli principle, with the following ansatz for the single-particle orbitals in a small volume $V$:
\begin{align}
 \psi_{\bf p}({\bf r}) = V^{-1/2}\left[ 1 + {\bf a}({\bf p}){\bf r} \right] e^{-\frac{i}{\hbar}{\bf p}{\bf r}}, \qquad |{\bf p}| \le p_F.
\label{eq:vw-ansatz}
\end{align}
where ${\bf a}({\bf p}){\bf r} \ll 1$ is assumed everywhere in $V$. The prefactor of the plane wave appeared to be the natural lowest order (in ${\bf r}$) choice to obtain gradient corrections to the local Thomas-Fermi result which is usually derived from pure plain waves. Using this ansatz von Weizs\"acker computed the density gradient 
\begin{align}
  {\bf \nabla} n= \frac{4}{\hbar^3}\int_0^{p_F} d^3p \, {({\bf r},t)\rm Re }\,{\bf a}(\bf p), 
\label{eq:vw-n-gradient}
\end{align}
 from which he obtained the kinetic energy $T[n;{\bf a}]$ of the system parametrically depending on ${\bf a}({\bf p})$. Minimization of the kinetic energy
yields a variational problem for the momentum dependence of ${\bf a}$, with the result [following, at least, for a power law ansatz ${\rm Re}\,{\bf a}({\bf p}) = a_l \cdot p^l)$] ${\bf a}({\bf p})=a_0={\rm const}$. Inserting this result in the kinetic energy straightforwardly recovers the result (\ref{eq:t}) with $\gamma=1$.

It is now interesting to make, again, contact with QHD. Indeed, using the result of this variational procedure gives, for the wave functions (\ref{eq:vw-ansatz}) and the density gradient (\ref{eq:vw-n-gradient}),
\begin{align}
 \psi_{\bf p}({\bf r}) &= V^{-1/2}\left[ 1 + a_0 r \right] e^{-\frac{i}{\hbar}{\bf p}{\bf r}} \equiv A_{\bf p}({\bf r})e^{-\frac{i}{\hbar }S_{\bf p}}, 
\qquad |{\bf p}| \le p_F.
\label{eq:vw-ansatz-result}
\\
  {\bf \nabla} n &= 2 \,n \,a_0.
\nonumber
\end{align}
This yields space-dependent amplitudes of the individual orbitals that are all identical, $A_{\bf p}({\bf r}) = A({\bf r}) = V^{-1/2}\left[ 1 + a_0 r \right]$, 
exactly as was assumed by Manfredi and Haas \cite{manfredi-haas01}, cf. Eq.~(\ref{eq:factorization}), in the derivation of the QHD equations.
Wave functions of the type (\ref{eq:vw-ansatz-result}) can, in principle, be realized for bosons in the ground state (all particles occupy the orbital 
with ${\bf p}=0$). At the same, time this ansatz does not yield the correct gradient correction of Kirzhnitz ($\gamma=1/9$).
Therefore, even though the variational minimization was performed correctly, the ansatz (\ref{eq:vw-ansatz}) turned out too simple an inadequate for a many-fermion system.

Von Weizs\"acker himself was well aware of the substantial deviations of his result from the experimental data \cite{weizsaecker}. He clearly pointed out the need to include correlation effects what led to the Coulomb corrections [as well as parity terms] in his famous formula for the total energy of nuclei. Since Coulomb corrections in nuclei are always small compared to the kinetic energy he correctly suspected that the problems lie in the Thomas-Fermi model itself. For applications to quantum plasmas or atoms this is true as well but, in addition, here Coulomb correlations play a much more prominent role, since they may easily be of the same order or even exceed the kinetic energy. Therefore, in the case of plasmas, use of the TF model (or of QHD) requires restriction to situations of weak coupling \cite{bonitz_pre_comm, bonitz_pscr}, cf. Sec.~\ref{s:qhd}.

\section{Conclusions and outlook}\label{s:summary}
In summary, we have presented an analysis of quantum hydrodynamic theory from the stand point of Thomas-Fermi theory. We have shown that the QHD equations have the correct structure in terms of powers of the density and lowest order density gradients. At the same time, the pre-factor of the Bohm potential in the QHD is an order of magnitude too large, and it depends on the system dimensionality. For 3D systems the proper choice that is consistent with the energy of the homogeneous Fermi gas is $\gamma=1/9$. The commonly used value, $\gamma=1$, applies to bosons in the condensate and is incompatible with fermions and the Pauli principle. Earlier QHD results for linear and nonlinear properties of fermions that are based on this value have to be revised \cite{dispersion}.

The present approach is straightforwardly extended to finite temperatures \cite{next} using the finite-temperature extension of TF-theory of Kirzhnitz \cite{kirzhnits75} and Perrot \cite{perrot79}. 
Finally, it remains to remove two major deficiencies of QHD: the neglect of exchange and of correlations. Here, the recourse to Thomas-Fermi theory with an exchange-correlation potential $U_{\rm XC}$ included in Eq.~(\ref{eq:e-funct}) gives a clear strategy how to proceed and to verify previous choices by Manfredi {\em et al.} \cite{manfredi08}.

\section*{Acknowledgements}
We  acknowledge useful discussions with L. Stanton and M.S. Murillo. MB acknowledges hospitality of Lawrence Livermore National Lab where this work was performed in June 2014.
This work is supported by the Deutsche Forschungsgemeinschaft via SFB-TR 24, project A5.

\end{document}